# *Moving beyond privacy and airspace safety: Guidelines for just drones in policing*

Mateusz Dolata ✉, dolata@ifi.uzh.ch, University of Zurich, Department of Informatics, Binzmühlestrasse 14, CH-8050 Zurich, Switzerland

Gerhard Schwabe, University of Zurich, Department of Informatics

## ABSTRACT

The use of drones offers police forces potential gains in efficiency and safety. However, their use may also harm public perception of the police if drones are refused. Therefore, police forces should consider the perception of bystanders and broader society to maximize drones' potential. This article examines the concerns expressed by members of the public during a field trial involving 52 test participants. Analysis of the group interviews suggests that their worries go beyond airspace safety and privacy, broadly discussed in existing literature and regulations. The interpretation of the results indicates that the perceived justice of drone use is a significant factor in acceptance. Leveraging the concept of organizational justice and data collected, we propose a catalogue of guidelines for just operation of drones to supplement the existing policy. We present the organizational justice perspective as a framework to integrate the concerns of the public and bystanders into legal work. Finally, we discuss the relevance of justice for the legitimacy of the police's actions and provide implications for research and practice.

# *Moving beyond privacy and airspace safety: Guidelines for just drones in policing*

## 1. INTRODUCTION

Drones may advance the public feeling of safety if they help to fight crime. Drones may reduce the public feeling of safety if they fly around without appropriate reason. While existing regulations address citizens' airspace safety and privacy concerns, police need guidance on when it is legitimate to use drones and how to design and use drones so that citizens, especially bystanders and witnesses of crime incidents, regard their usage as legitimate. This paper addresses the police's needs by developing guidelines for just drones. Prior studies on police drone usage surveyed citizens to gather their general opinions; we used design science research methods to a) propose a drone usage scenario b) expose citizens as bystanders to a drone trial, c) gather their situated opinions, d) systematically analyze their feedback, and e) craft guidelines for just drone usage.

Drones have grown in popularity and are applied in many scenarios (Çetin et al., 2022; Sabino et al., 2022). The police use them for search and rescue, surveillance, and tracking suspects (Liu et al., 2022). Studies have explored the public's attitude toward drones in general (Sabino et al., 2022) and, to a lesser extent, with a focus on law enforcement (Heen et al., 2018; Sakiyama et al., 2017). However, the research still lacks insights on when public accepts drones as a *legitimate* mean to be used in law enforcement. The existing regulation of drones is lagging behind the technological progress, focuses on safety and privacy leaving other aspects of drones' use unattended, lacks comprehensive empirical grounding, and ignores various relevant social groups including witnesses and bystanders of an incident (Bracken-Roche, 2018; Fox, 2022; L. K. Johnson et al., 2017; Smith, 2015). This manuscript discusses the results of a field study designed to inform policy development for use of drones in law enforcement, which indicate that bystanders are in a dilemma regarding the justice and advantages of police drones. We interpret the findings in the context of organizational justice theory (Greenberg, 1986) and develop a catalogue of guidelines to inform policymaking.

Drones, like many other technologies, enhance the capabilities of the police. They also change the balance of power between the citizens and law enforcement agencies. The regulatory landscape remains dynamic and inconsistent, and many stakeholders raise concerns and provide rationales for or against specific protocols and police practices (Fox, 2022; Smith, 2015). Like with other novel technologies (Dolata & Schwabe, 2023a), the public is currently in the phase of negotiation and interpretative flexibility. This opens the opportunity to study and influence how citizens react to drones in police operations and how those attitudes should be reflected in the regulatory framework. This requires a cautious and considerate engagement with various groups of citizens and their thoughts.

Existing regulations primarily focus on airspace safety and privacy, with drone laws being fragmented and inconsistent across regions (Smith, 2015). These regulations lag behind those for smart grids or autonomous cars (Lee et al., 2022), often lacking empirical basis and potentially influenced by specific interest groups (PytlikZillig et al., 2018; West et al., 2019). Stakeholder involvement in policy development, including citizens, is inadequate, leading to claims of insufficient democratic support (L. K. Johnson et al., 2017). The needs and attitudes of bystanders, those directly affected by a crime without their involvement, are rarely discussed. If mentioned, it's usually in the context of physical safety as potential victims of insecure drone maneuvers (AL-Dosari & Fetais, 2023; Altawy & Youssef, 2016; Clothier et al., 2015), not as participants in a complex operation. This is surprising given the police's reliance on bystander cooperation for safety and conviction (Tyler & Fagan, 2008). Most research focuses on the general public's perspective without considering their exposure to drones in specific scenarios (e.g., Heen et al., 2018), with few exceptions addressing general drone use rather than law enforcement (e.g., Miron et al., 2023). Despite drone policy development beginning in the late 2000s, significant shortcomings remain. This leaves police and politicians uncertain about drone use for law enforcement and the necessary steps to ensure legitimacy and cooperation from individuals on-site.

Our goal is to establish empirically informed propositions that can be used to supplement the existing regulations and responds to the problem of incomplete policy and lack of empirical grounding for it. The manuscript relies on a design science research (DSR) project. DSR offers a reliable guidance and rigorous approach for empirically grounding, designing, and evaluating technical and non-technical artefacts. DSR also provides guidance how to rigorously tie empirical, conceptual and theoretical insights together with creative approaches to design and evaluate a solution artefact (Dolata & Schwabe, 2016). This makes it increasingly popular as a scientific method for the development of various sorts of artefacts including guidelines and rules (Offermann et al., 2010; Weigand et al., 2021). We build on this tradition to propose a catalogue of guidelines for use of drones by the police.

The study outlines twelve categories of concerns about technology, operation, society, and regulation that affect citizens' sense of safety and acceptance of drones. This data interpretation provides a framework for understanding these concerns through the lens of organizational justice. It also offers a set of guidelines related to procedural, distributive, and interactional justice. Ensuring justice and fair treatment is deemed a core police task (CED, 2023; Fox, 2018) and is crucial for the police's societal legitimacy (Fielder & Murphy, 2022; Tyler, 2021). We deduce that drone usage should promote fair treatment of individuals, along with enhancing efficiency and safety.

## 2. BACKGROUND

Drones' usage in various settings has garnered considerable attention in literature. Technical research aims to enhance drone capabilities, while social research examines public attitudes towards drones. A broader perspective offers comprehensive analysis on factors influencing public acceptance of drones in different usage contexts. Some studies specifically explore police drone use and public reactions. Legal research tackles regulatory aspects of drones and policing. These perspectives are crucial for an informed discourse on the legitimacy of police drone use.

### 2.1.   Drones

Multiple terms are used about what the public calls a drone (Fox, 2022). Most generally, the terms *unmanned aerial vehicle* (UAV) and, simply, *drone* are used. However, aviation agencies mainly refer to *remotely piloted aircraft* (ICAO, 2017). Some legislations use *unmanned aircraft*, whereas many US researchers prefer the unmanned aerial system phrase (Aydin, 2019; Sabino et al., 2022). Since '*drone'* is broadly embraced and understandable across disciplines and languages, we use this term.

The term "drone" has various definitions. Originally, it referred to a a stingless male bee (MW, 2023). Now, it describes an airborne device, either autonomous or remotely piloted (Fox, 2022). The term covers machines from insect-sized crafts to large airplanes akin to commercial jets (Çetin et al., 2022, p. 1). Typically, it refers to equipment under 20 kg, powered by rotary, fixed, or flapping wings, ducted fans, or tilt-rotors (Garnica-Peña & Alcántara-Ayala, 2021). Drones with multiple rotary wings are known as multicopters (Stampa et al., 2021). Micro drones, weighing under five kg, with a range of about ten kilometers, an endurance of one hour, and a maximum flight altitude of 500 m, are the most common (Garnica-Peña & Alcántara-Ayala, 2021). In this article, "drone" refers to micro multicopters.

Forecasts predict a steady growth in the drone industry for individual, commercial, and governmental activities, with a predicted impact of over EUR 10 billion and 100,000 jobs created in Europe by 2035 (Çetin et al., 2022; EU, 2017). The global drone market was valued at USD 22.5 billion in 2020 and is expected to double by 2027 (Sabino et al., 2022). A fleet of 400,000 drones is likely to be used in Europe by 2050 (EU, 2017), indicating the rising prevalence of drones, particularly in commercial and governmental applications. This makes the study of drones increasingly important.

## 2.2.  Acceptance of drones

Nonetheless, the proliferation of drones is subject to limitations related to (1) societal acceptance and (2) regulations. Acceptance of drones attracted interest from social and aviation sciences yielding a wealth of studies (e.g., Ahrendt, 2020; Boucher, 2016; Chang et al., 2017; Lin Tan et al., 2021; Miron et al., 2023) and meta-reviews (e.g., AL-Dosari et al., 2023; Altawy & Youssef, 2016; Herdel et al., 2022; Legere, 2019; Sabino et al., 2022; Tepylo et al., 2023). Given the interest sparked by this topic and the wealth of individual studies, we rely on the literature reviews for the understanding of current results concerning the general attitude towards drones. Even though many studies offer a multi-dimensional perspective combining characteristics of the respondents as well as the features of technology or its use, we focus on those aspects that can effectively inform the development of a potential policy.

Past results suggest that drones are more widely accepted for public safety than for commercial, recreational, or military use (Aydin, 2019; Boucher, 2016; Eißfeldt et al., 2020). The public generally supports drone use in disaster response, search and rescue, research, infrastructure monitoring, and medical transport (Eißfeldt et al., 2020; Lin Tan et al., 2021; Sabino et al., 2022). Public agencies, including police, are the most preferred drone operators, followed by commercial entities, and lastly, private users (Ahrendt, 2020; Lin Tan et al., 2021). However, a consistent pattern across studies shows that about 40% of respondents have negative views on drone use in any context (Eißfeldt et al., 2020). The literature suggests that the scenario in which drones are used greatly influences public acceptance, with some uses seen as more legitimate than others.

A significant part of drone acceptance research deals with the risks people ascribe to using drones. Çetin et al. (2022) and Sabino et al. (2022) review the existing literature and identify the following concerns: (1) misuse, theft, and crime; (2) malfunction and property damage; (3) technical safety and cybersecurity; (4) emissions, noise, and visual pollution; (5) fear, negative emotions, lack of human interaction, and missing communication; (6) legal and regulatory issues; (7) impact on work processes and job market; (8) privacy and data protection. Overall, if the acceptance of drones is to improve, drone providers and regulators must give attention to these risks.

Studies link acceptance to various population features, including demographic characteristics (Del-Real & Díaz-Fernández, 2021; Eißfeldt et al., 2020; Meiländer, 2021), public knowledge or attitude towards technology (Aydin, 2019; Eißfeldt et al., 2020; Lin Tan et al., 2021), and previous drone exposure (Kähler et al., 2022; Miron et al., 2023). Despite the relevance of individual studies, the sample and context differences make it challenging to find consistent patterns (Sabino et al., 2022). Our focus is on designable and regulatable aspects rather than individual characteristics. We encourage readers to refer to summaries and studies for a more specific discussion on the impact of demographic background, political preferences, or individual technological affinity on drone acceptance (e.g., Anania et al., 2019; Eißfeldt et al., 2020; Lidynia et al., 2017; Rice et al., 2018; Sabino et al., 2022; Tepylo et al., 2023). Given the numerous studies on individual differences in drone perspectives, we agree with Boucher (2016, p. 1391) who suggests that research should focus on "developing drones acceptable to citizens" rather than analyzing citizens' acceptance of drones. We propose a catalogue of guidelines for drone application in law enforcement, following this argument.

While numerous acceptance studies offer valuable insights, there is room for improvement. Many studies distinguish individuals based on inherent traits like age or gender but overlook the role a person plays in a specific scenario. There is little attention to the degree of a person's involvement or reliance on drone operations. Attitudes may vary if a person is unaffected by, witnesses, or depends on a drone, especially if they're lost or injured in an inaccessible area. This makes it challenging to create policy that addresses the conflicting needs of these groups. The most relevant studies only outline potential use contexts (Herdel et al., 2021; Lidynia et al., 2017; S. R. Winter et al., 2016) and target those previously exposed to drones (Miron et al., 2023), or who were introduced to the technology outside of an operational context (Chang et al., 2017). The stimuli used may not offer enough context for subjects to envision themselves in a real situation. Many studies use decontextualized survey questions, textual stimuli ("Imagine a situation where your local police department announce plans to use UAVs...", cf. Winter et al. (2016, p. 129)), or pictures or videos (Ahrendt, 2020; Kähler et al., 2022; Oltvoort et al., 2019; PytlikZillig et al., 2018; Tolmeijer et al., 2022). However, it is unclear to what extent subjects can form a reliable basis for assessment.

While past literature offers insightful results, we question their applicability in understanding the public's specific reactions during an incident. Many agencies prioritize these reactions to improve the sense of safety among witnesses or victims, rather than causing further fear. Law enforcement agencies also aim to ensure the public perceives drones as legitimate, encouraging them to follow drone guidance and cooperate with police using drones. We suggest that a drone can be subject to legitimacy attributions if it is granted the right to enforce rules or is perceived to have the authority to make decisions (cf. Stillman, 1974). This is crucial in establishing policies that not only address general drone acceptance but also ensure individuals react predictably in situations involving drones.

Collecting opinions from bystanders during actual drone operations, particularly in emergency situations like crime incidents or search and rescue, is challenging. Bystanders may be traumatized, and without appropriate policies, public agencies may hesitate to use drones. Additionally, some witnesses may decline to share their experiences, potentially leading to self-selection bias in research. Given these factors, our data is sourced from a field trial conducted during police training. We examine attitudes, concerns, and views from participants acting as bystanders, witnesses, and professional police officers in a simulated bank robbery scenario.

## 2.3.    Drones in policing

The use of drones for policing is a growing area. The first reported drone operation in prosecution occurred in 2011 when a drone assisted a North Dakota sheriff's search for three armed men on a parcel of 12 square kilometers (Fox, 2018). According to a 2020 survey, 1'000 US police agencies already employ drones in their work (Gettinger, 2020; Liu et al., 2022). In Switzerland, the police in 13 out of 26 cantons possessed drones as of 2017 (Klauser, 2021b). It is expected that the use of drones in policing will grow to 50'000 devices to be used in 2050 in Europe (EU, 2017). This trend will likely accelerate with the increasing autonomy of drones and cheaper technology. Research in policing and criminal justice literature explores the use of drones by the police taking police's and citizens' perspective indicating what advantages are related to the use of drones and what are the general opinions about it. Yet, there is lack of studies which rely on citizens' exposure to police's actions involving drones.

A policy for drones' use by the police must be subject to the mandate of the police. Police's tasks embrace ensuring "fairness and legality in an area of public life" and preserving "law and order" (CED, 2023; Fox, 2018). The frequently employed *crime control* model assumes that the police needs to efficiently "apprehend, try, convict, and dispose (...) a high proportion of criminal offenders whose offenses become known" in a context of limited resources and pressure for speedy, ultimate, and hard to challenge prosecution effects (Liu et al., 2022; Packer, 1964, p. 10). This pressure on public safety agencies has increased (Stampa et al., 2021). Drones are seen as means to make the police's work more efficient and less risky, e.g., by reducing the presence of police officers (Anania et al., 2019; Liu et al., 2022; Sabino et al., 2022). Swiss police officers who already use drones see the development of "novel services," cost savings, and risk reduction as the main advantages (Klauser, 2021a). However, it is controversial if drones use contributes to the quality and reliability of law enforcement postulated by an alternative perspective on police's work, i.e., the *due process* model (Liu et al., 2022). This tension has implications for the acceptance of police drones.

Survey-based research shows that the use case significantly influences public perception of drones. Many US citizens view drones as valid tools for search and rescue, highway accident surveillance, and tactical operations during bomb threats or active shootings (Heen et al., 2018; Sakiyama et al., 2017). About 70% of respondents supported drone use for tracking fleeing suspects, ranking it fifth out of ten use cases (Heen et al., 2018). Canadians endorse police drone use for disaster response, locating missing persons, handling hostage situations, and emergency response (Saulnier & Thompson, 2016). However, drone use for detecting traffic violations, issuing speeding tickets, identifying individuals at sports events, routine patrols, monitoring political protests, and supervising house arrest offenders is contentious (Heen et al., 2018; Saulnier & Thompson, 2016). Drones have been used to enforce COVID-19 restrictions (Liu et al., 2022; Mora et al., 2021; Sarkar, 2021). Existing research suggests the public is more accepting of reactive drone use, such as investigating a committed crime or genuine suspicion, than proactive use for monitoring and crime prevention (Heen et al., 2018; Sakiyama et al., 2017). The police can use drones in many scenarios (Liu et al., 2022), but not all potential uses have been empirically studied for public acceptance.

Studies on drone use in law enforcement echo concerns from general drone research, such as privacy, and add context-specific findings (e.g., privacy). Citizens prioritizing public safety, identifying as conservative/authoritarian, and living in racially diverse US neighborhoods, tend to support drone use more (Anania et al., 2019; Heen et al., 2018; West et al., 2019). Factors like past police interactions, knowledge of police drone use, personal characteristics, and expected drone use outcomes influence drone acceptance in policing (Sakiyama, 2017). These findings reflect those in generic studies (see 2.2). They focus on the the individual's social and political standing.

The literature attends to the concerns raised by the public. Privacy is the most frequently postulated and researched concern (Bentley, 2018; Liu et al., 2022; Nelson et al., 2019; Nelson & Gorichanaz, 2019; Sakiyama, 2017). Independent of their nationality, European citizens decline to sacrifice their privacy for security (Gómez et al., 2015; Miron et al., 2023). Further concerns pertain to the risk that drones might disproportionally target vulnerable communities and people with unpopular, opposing views and strengthen the surveillance tendencies in the state (Liu et al., 2022). While those considerations seem appropriate in light of fairness debate around law enforcement (Council et al., 2004), they give little guidance on how drones should be employed by the police to bridge the security needs and the potential risks related to the use of drones. Further, whereas assuring privacy and safety might be the necessary condition, it remains open whether it is sufficient to make people see drones as a legitimate mean in law enforcement and support overall police's legitimacy to exercise power.

## 2.4. Regulation of drones

The literature agrees that drones should be subject to legal oversight (Fox, 2022; Heen et al., 2018; Sabino et al., 2022; Sakiyama et al., 2017). The regulation is seen as a way to balance the conflicting interests of citizens, states, and commercial drone users (Bentley, 2018; Fox, 2017, 2022). Simultaneously, researchers and legal experts agree that the existing legislation did not keep up with technological developments in pace and scope (Fox, 2019; Shenoy & Tyagi, 2022). Many laws seem detached from the empirical basis and too abstract to be applicable (Nelson et al., 2019; Nelson & Gorichanaz, 2019; Smith, 2015). Smith (2015, p. 423) claims they "provide only the illusion of regulation." Finally, researchers question whether the policymaking processes in this area consider public opinion and citizens' concerns or, instead, follow extreme communities' ideological and personal preferences (PytlikZillig et al., 2018; West et al., 2019). Even though issues like trust have been identified relevant for the development of adequate policy for the use of drones (Nelson & Gorichanaz, 2019), the implementation of those insights in policy remains lacking. Issuing adequate regulation is a challenge.

Two central drone aspects are subject to existing regulation: (1) civil airspace safety and (2) drone operation. US Federal Aviation Administration issued an act on the Operation and Certification of Small Unmanned Aircraft Systems which lists formal requirements for drone pilots or operators and necessary physical limitations, including information about the allowed flight altitude (Bentley, 2018). However, usage-related regulations at the national level failed, leading to state-specific laws, of which only some consider public concerns (Bentley, 2018). Many states prohibit the warrantless use of police drones in places where privacy can be expected unless they are used for preventing an immediate danger, search for an escaped prisoner. The US states also differ in how they treat evidence collected in the warrantless usage of drones and rule concerning data archiving (Bentley, 2018).

In Europe, drone-related regulations are also fragmented. There are rules addressing the safety risks covered in standards issued by the European Union Aviation Safety Agency, applicable in most European countries in and outside of the EU (Fox, 2019). Further, EU bodies declared the initiative to address further concerns (Fox, 2018). The core of the European drone policy is included in the Aviation Strategy for Europe from 2015 and subsequent protocols (EU, 2015, 2019a, 2019b; Fox, 2022). The regulations mention "public acceptance" as the key to the proliferation of drones. In this regard, the legislation explicitly attends to (1) citizens' fundamental rights, including privacy and data protection, (2) nuisances such as noise, (3) security risks related to malicious attacks and uses, (4) identifiability and accountability of the operator through, e.g., a registration platform or an identification chip (Fox, 2022). EU member states are expected to address those aspects in their national laws.

Despite the efforts to establish a comprehensive framework for drone operations, the issued regulations have not yet achieved stability. While the airspace safety rules are operationalized and specific, the rights concerning the drone's use and public perception remain abstract and declarative (Smith, 2015). They focus on privacy issues, accountability, and security, whereas other concerns, such as communication, human-drone interaction, or adequacy, remain unanswered (Bentley, 2018). Finally, many regulations abstract from the purpose of drones' use such that, e.g., the limitations concerning the altitude or certification requirements are the same for law enforcement agencies and commercial or private users. Clearly, the policymaking process concerning drones is not yet finished, and more efforts will be necessary to establish sustainable standards.

Regulatory work strengths and weaknesses are evident in police assessments of drone rules. Safety and security regulations are generally accepted, with 78% and 84% of Swiss police officers, surveyed by Klauser (2021a), agreeing that drones should avoid flying near crowds or dangerous areas. Additionally, 83% believe drones should only be flown within the pilot's sight, and 94% support mandatory insurance for operators. However, the current legal framework is viewed as inadequate. Only 38% of officers believe current drone laws are sufficient to combat terrorism, 49% feel they prevent accidents, and 39% consider them adequate for privacy protection. These figures are low compared to private companies' assessment, which was 49% for terrorism prevention, 71% for accident prevention, and 78% for privacy protection. Overall, there is consensus that regulations need improvement. Greater attention should be paid to citizens' concerns and needs. We argue that a detailed exploration of citizens' reservations will foster a well-informed, sustainable policy that enhances police efficiency, improves quality, and ensures police legitimacy.

## 2.5.    Relevance of justice for police work

Police require legitimacy, generally defined as the right to exercise power (Tankebe et al., 2016), to function. The normative discourse about legitimacy deals primarily with questions on the roots of an institution's right to exercise power (Adams, 2018). Descriptive positions assume that an institution's legitimacy comes down to the beliefs or faith of the society about an institution or a person, which are "lent prestige" to exercise power (Mommsen, 1992; Peter, 2017; Weber, 1964). This perspective is particularly important for the day-to-day work of state agencies like the police (Worden & McLean, 2017). To operate in today's complex environment, police rely on cooperation with the public. They must ensure that the citizens trust and support them, i.e., that they attribute legitimacy to the police.

The use of technology was shown to impact the perception of the police as a trustworthy and legitimate law enforcement agency (Bradford et al., 2020; Davies, 2022; Demir et al., 2020; Heen et al., 2018; Mrozla, 2021). If the public trusts the police, citizens are more likely to approve of using novel technologies (Bradford et al., 2020). There is a circular relationship between the use of technology, the police's legitimacy, and the acceptance of new technologies.

Legitimacy, trust, and willingness to cooperate are established topics in criminal justice (Fielder & Murphy, 2022; Paternoster et al., 1997; Tankebe, 2013; Tyler, 2021). Victimization research indicates that victims who experienced the treatment by the police as just were more likely to see police's action as legitimate and support them (Paternoster et al., 1997; Wolfe et al., 2016). The connection between justice and legitimacy, trust, and cooperativeness was shown to hold for the general public as well (Bolger & Walters, 2019; Council et al., 2004; Davies, 2022; Sunshine & Tyler, 2003), but the effect might be weaker for marginalized groups or immigrants (Bradford & Jackson, 2018).

The aspect of justice demands careful consideration. Justice is a complex concept and has been assigned many meanings. One of the oldest definitions is attributed to Roman Emperor Justinian: "justice is the constant and perpetual will to render to each his due" (Miller, 2003, p. 76). It indicates that justice pertains to the rights of an individual as opposed to, e.g., the prosperity of a nation (Miller, 2003). The issue of justice emerges when various people raise conflicting claims or demand the realization of specific rights, e.g., to freedom, privacy, security, and good treatment (Miller, 2021). Justice is different from fairness. Fairness refers to impartiality, meaning adopting a neutral standpoint independent of one's special desires, biases, or preferences (Cottingham, 1983; Velasquez et al., 1990). Even though *justice* and *fairness* are increasingly used interchangeably, the differentiation might be useful to discriminate between technologies that assure fairness, algorithmic fairness (Dolata et al., 2022), and technologies' impact on justice or perceptions thereof.

Research frequently attends to perceptions of justice implicitly assuming that assessments and notions of justice are, at least to a certain extent, constructed by an individual or in socio-cognitive sense-making processes (Beugré, 2009; Degoey, 2000; Dolata & Schwabe, 2023b; Lamertz, 2002; Lind et al., 1998). A family of justice theories tries to capture and identify characteristics that contribute towards individuals' and groups' assessments of an individual, an organization, or an action as more or less just (Masterson & Tong, 2015). The literature on justice often suggests that a person's perception of justice is primarily based on their experience of a specific event (R. Cropanzano et al., 2015; Masterson & Tong, 2015). For instance, when police prosecute an individual, this person as well as the society form their perception of justice related to this situation based on various aspects: distribution (*Who receives what? Is the outcome or distribution of penalty or reward adequate?*), interaction (*Is the treatment or benefit adequate to the suspect's input or behavior in terms of dignity and respect? Is the information provided timely and sufficient?*), and procedure (*How was the decision taken? Was the procedure impartial and free of bias?*) (Colquitt et al., 2013). In other words, distributive justice deals with allocating resources and positive and negative outcomes among the affected individuals (Colquitt et al., 2005, 2013). Interactional justice stresses that individuals expect truthful and respectful treatment if they act accordingly (Bies, 2001; Dolata et al., 2022). Interactional justice can be divided into interpersonal and informational components. Interpersonal justice reflects "the degree of respect and propriety," while informational justice denotes "the degree of justification and truthfulness" during procedures (Colquitt & Rodell, 2011, p. 1183). Finally, procedural justice demands justification so affected individuals can voice their concerns (Greenberg, 1986).

Feinberg (1974) initiated decades of research by distinguishing between distribution and procedure in justice. Bies and Moag (1986) later added interaction and communication as factors influencing justice perceptions (Masterson & Tong, 2015). Greenberg (1986) unified these aspects into the concept of organizational justice, which initially focused on employee perceptions of their organization and superiors. The theory's scope has since broadened to include justice related to technology (Dolata et al., 2022; Robert et al., 2020), client-provider relationships (Blodgett et al., 1997; Choi et al., 2016), and social context (Gollwitzer & van Prooijen, 2016; Masterson & Tong, 2015). It now examines perceptions of justice by individuals or groups towards organizations or power-holding authorities. Justice perceptions influence trust in authority figures (Colquitt & Rodell, 2011). Therefore, we primarily use organizational justice theory for its holistic approach to justice dimensions.

Research has analyzed how perceptions of police justice affect attitudes towards the police. Key elements of procedural justice in policing include fair process, transparency, voice opportunity, and impartiality (Schaap & Saarikkomäki, 2022; Tyler, 2021)[1]. Studies have shown that procedural, interactional, and distributive justice can boost public confidence and willingness to cooperate with the police (Aston et al., 2021). Police behavior and actions, such as treating citizens with dignity and respect (McKee et al., 2022; Skinns et al., 2020; Wemmers et al., 1995), equitable resource allocation, law enforcement practices, and outcomes (Tankebe et al., 2016), and adherence to the law (Tankebe, 2013) all influence perceptions of police legitimacy. A citizen's overall sense of justice can also enhance police legitimacy (Mazerolle et al., 2013).

---

[1] Tyler's and Greenberg's understanding of procedural justice overlap but are not identical. They are connected concerning their origin in Feinberg' work. However, their conceptualizations have drifted apart. Generally, Tyler's procedural justice is a broader term which captures aspects of interactional justice as well. In this article, whenever we refer to procedural justice, we mean the notion included in organizational justice, unless it is clearly specified otherwise (e.g., *Tyler's procedural justice*).

However, the relationship between technology use by the police and the perception of justice remains underexplored. Some literature indicates that body-worn cameras enhance police's procedural justice and accountability, improving perceived legitimacy (Davies, 2022; Demir et al., 2020) leading to enhanced acceptance (Bromberg et al., 2020). The relation is mutual: perceptions of procedural and distributive justice impact citizens' attitudes toward using technology by the police (Mrozla, 2021). However, some technologies might negatively impact perceived legitimacy (Rossler, 2019). A more considerate view of the relationship between technology and justice is necessary (Schaap & Saarikomäki, 2022). We claim that the presence of technology does not in itself affect justice. We must explore what aspects of technology, or its operation, generate what response from the public. We must learn how to build and operate technology to be just and perceived as such. Only then the use of technology during law enforcement be sustainable. Using the notion of organizational justice, this study explores the relationship between perceptions of justice, acceptance of drones, and police duties.

## 3. METHOD

This study is part of a larger, ongoing project with a Swiss police department (see Figure 4). The collaboration primarily aims to comprehend when and how police should utilize drones. Furthermore, the project aims to develop suitable drones, technical oversight, control mechanisms, and operational procedures. It also focuses on creating effective organizational processes, structures, role distribution, and policy propositions to facilitate the efficient use of drones in alignment with police objectives and societal rights. For this purpose, the authors are collaborating with political philosophy researchers. Ultimately, the project will produce an artifact network (Weigand et al., 2021), leading to the development of an effective sociotechnical system where drones assist police in law enforcement operations.

DSR is a research paradigm aimed at enhancing human knowledge and practice through the design of innovative artefacts (Hevner et al., 2004). According to the three-cycle view of DSR (Hevner, 2007), the proposed guidelines are seen as the design artefact. The field trial and police input provide the domain knowledge needed for the relevance cycle of DSR. Literature on acceptance and drone-related regulations offers grounding for the rigor cycle. The guidelines are part of a larger artefact network (Weigand et al., 2021), which includes drones, their software, police workflows, hierarchical structures, etc. Thus, this study is part of a larger project with interrelated goals, the description of which exceeds the scope of a single manuscript. Despite DSR's predominant use for the development of technical artefacts, many researchers use DSR to design non-technical artefacts (Offermann et al., 2010; Weigand et al., 2021) including rules (Parsons & Wand, 2008), recommendations (Pries-Heje & Baskerville, 2006), constructs (Goldkuhl & Karlsson, 2020), and guidelines (Weigand & Johannesson, 2023). We follow this tradition proposing a catalogue of guidelines for operation of drones in law enforcement.

In July 2021, researchers conducted a field trial to identify suitable conditions for drone usage. This trial investigated the use of drones in capturing a bank robber and assessing bystander reactions [2]. The trial was conducted at a police training facility, replicating a real-life operation four times over two days with diverse participants. The operation procedures mirrored standard police practices. Trained police officers piloted the drones, which were police property. However, new elements were introduced to better integrate the drones. The drone's video signal was made accessible to the command center, and to emulate autonomous flight, pilots adhered to a strict protocol. This trial provided a shared experience for all participants, which was referred to in subsequent data collection. Post-trial, group discussions were recorded and used as the basis for our findings. Further details follow.

## 3.1.    Technical and scientific knowledge base

Before the field trial, the research team conducted preliminary activities in 2020. This included integrating the research with the knowledge base, following Hevner's (2007) rigor cycle. Additionally, we examined the drones' technical capabilities based on hypothetical scenarios outlined at the project's inception, adhering to Hevner's (2007) relevance cycle.

We initiated the exploration of the knowledge base with a narrative review in 2020, highlighting regulatory challenges of police drone usage (cf. Figure 2). We continually added relevant literature to this base throughout the project, eventually encompassing over 150 drone-related articles. Given the vast research in this area, we selectively focused on specific literature to maintain a manageable overview. We utilized numerous reviews for general drone research and their use in emergency services (e.g., AL-Dosari et al., 2023; Altawy & Youssef, 2016; Herdel et al., 2022; Legere, 2019; Sabino et al., 2022; Tepylo et al., 2023). We concentrated on individual studies that touched on policy aspects of police drone work. The outcomes of this process are detailed in Sections 2.2 and 2.3.

---

[2] By "realistic field trial," we refer to a situation that could occur in real life, enacted under safe conditions for all involved, including police. The trial primarily tested the use of drones and bystander perceptions. We recognize the trial's confined nature may influence participant perceptions. However, we've made significant efforts to ensure maximum realism. For example, police intervention and patrol units were genuinely unseen by participants until necessary. Conducting the trial in a city could create unnecessary outrage and risk traumatizing uninvolved individuals. Given the limited use of drones in actual law enforcement operations like this, there's hardly any chance to interview real witnesses. Such interviews also raise ethical issues due to potential trauma and practical concerns due to limited access.

To comprehend the potential of drones in police work, we joined the police that recently purchased several drones and trained two pilots. We investigated the technical capabilities of drones in a controlled environment, testing the speed and reliability of an autonomous drone moving from a base to a crime scene and tracking a subject. The results informed the design of the field trial (see Section 3.2).

## 3.2.    Field trial scenario

The scenario for the field trial was proposed by the police and further developed in collaboration with the authors, a student assistant, and two doctoral students supervised by the authors. The specification rooted in a preliminary study of literature and technical trials to identify useful abilities of autonomous drones (as of 2020) and test them in confined conditions without risks for human subjects. The authors, two designated police representatives, and the student assistant were meeting at least once every month between the beginning of 2020 and July 2021 to specify the scope and procedure of the field trial, so it was aligned with the project partners' interests while obeying the regulations and research practice.

The trial scenario revolved around a bank heist, with an actor assigned to the role of the robber. This individual's primary task was to trigger the alarm and make a swift escape. They were allowed to utilize any strategy they saw fit to conceal themselves in the area, making it hard for the police to find and capture them. Remarkably, in two of four sessions, the robbers managed to flee the training premises, heightening the realism of the situation for the study subjects and police.

The scenario involved 52 subjects who acted as persons who, by chance, happened to be near the crime scene or witnessed the robber fleeing. We refer to them as bystanders. They were informed that they participated in a trial testing a novel intervention approach and that police would be involved. They did not know about the scenario or the drones and were not familiar with the area in which the trial was set. Bystanders were positioned in an area designated as the "town square" and received micro-tasks unrelated to the robbery. For instance, two of them had to determine if they shared the same friends or hobbies. The idea was to improve the realism of the situation and mimic the attention level they would likely have when being out in a city. During the sessions, bystanders were allowed to act freely. They could react to the police's requests or the robber's behavior as they saw fit. Figure 1 provides an overview of the used training facility.

In addition to bystanders and the robber, the scenario involved police officers. There were two police troops: an armed patrol and an intervention unit. Armed patrols secure the crime scene, get uninvolved individuals to safety, deliver helpful information to the command center, and prepare the environment for the intervention. Intervention units search the area and try to convict the offender. The troops' operation was coordinated from a command center in the police headquarters approx. 20 km away from the facility. Two police officers were overseeing the operation from the command center. The command center had access to the video stream from the drone and was communicating with the troops via radio. All police officers were informed that the operation involved a drone, and their participation in the field trial was framed as a training session.

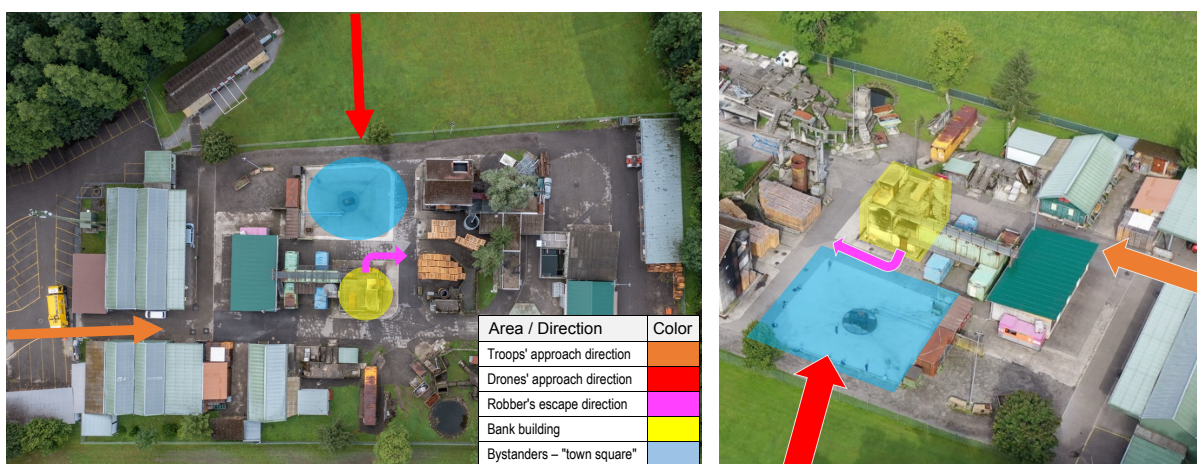

***Figure 1.*** *The training facility used for the field trial with indicated relevant areas.*

While the bystanders were in their tasks, the robber triggered an alarm which started the overall action. The command center launched the operation by quickly ordering the drone and police troops to be dispatched 20 seconds after the alarm. The drone left its base 800 m away and followed the robber, who had already left the bank. After one minute after the alarm, the drone was up in the air. The armed patrol and intervention unit arrived in 3 and 5 minutes, respectively. The time to catch the robber was between two and twenty minutes. The operation ended after the police caught the robber.

The sequence implemented is based on police analysis of previous incidents, demonstrating that a drone can reach a crime scene about twice as fast as the closest patrol unit. Initially, the drone is the sole police representative on site, raising questions about its legitimacy and public perception. The drone was consistently used to aid the operation. Based on prior trials and the need for superior technical capabilities, all tests were conducted using the DJI Matrice M210 V2 Drone, equipped with a zoom camera (DJI Z30) and a thermal imaging camera (DJI XT2). Trained police officers piloted the drones, adhering to a strict protocol (arrive at the scene, identify the robber using a description, track the robber). This simulates the recognition and tracking abilities of autonomous drones identified in earlier trials. However, due to security and regulatory constraints, automatic features were not used with uninformed test subjects, hence relying on the pilots' expertise.

### 3.3.   Participants and data collection

We employed non-random convenience sampling with the aim to obtain variety of opinion and statements in group discussions. The bystanders were recruited through a commercial recruiting platform and a university platform. We asked for about an equal distribution of age and gender, and we indicated variety as a desired feature. The study participants had to speak German at a communicative level. For their participation of overall 2 hours, the participants were reimbursed CHF 70 according to the University's practice (25 francs per hour plus travel costs). The two hours included preparation, demographic data collection, trial participation, subsequent group interviews, and scenario resolution.

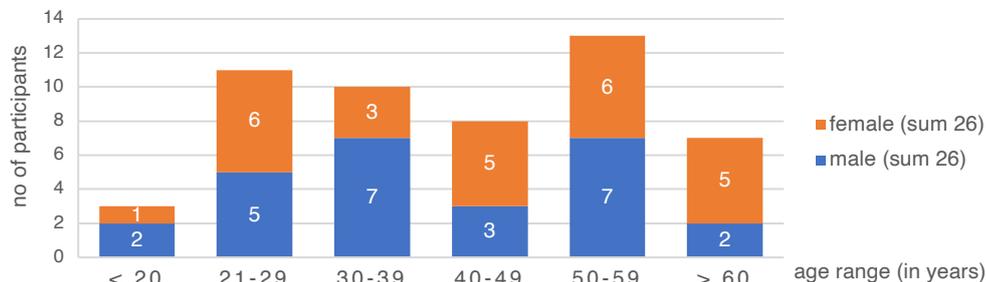

***Figure 2.*** *Age and gender distribution among study participants.*

Overall, 52 participants were recruited. 26 identified themselves as men and 26 as women. The participants' ages were nearly equally distributed across the specified age ranges (cf. Figure 2). 25 (48%) participants came from a city, while the rest were from suburbs or the countryside. This deviates slightly from the Swiss regional distribution, where 63% of participants live in core urban areas (BAS, 2023). The participants varied in terms of achieved educational degrees, however this also deviates from distribution in Switzerland (cf. Figure 3). Overall, the participants formed a diverse population with distribution corresponding to Swiss society in terms of age and gender.

The participants were equally assigned to one of four trial sessions conducted on two consecutive days in July 2021. After the sessions, the participants attended a group discussion moderated by a research team member. The discussions were semi-structured according to eight topics: the overall impression, feeling of safety, feeling of being disturbed, the perception of drones, the accountability of drones, the responsibility related to a drone's actions, drones' applicability, and reasonability. The participants were free to express their opinions and engage in discussions. If necessary, the researchers ensured that all subjects had a chance to share their opinion. The discussions lasted up to 45 minutes and were in (Swiss) German. The groups' sizes varied from four to seven participants. Overall, ten group discussions were conducted, and audio recorded.

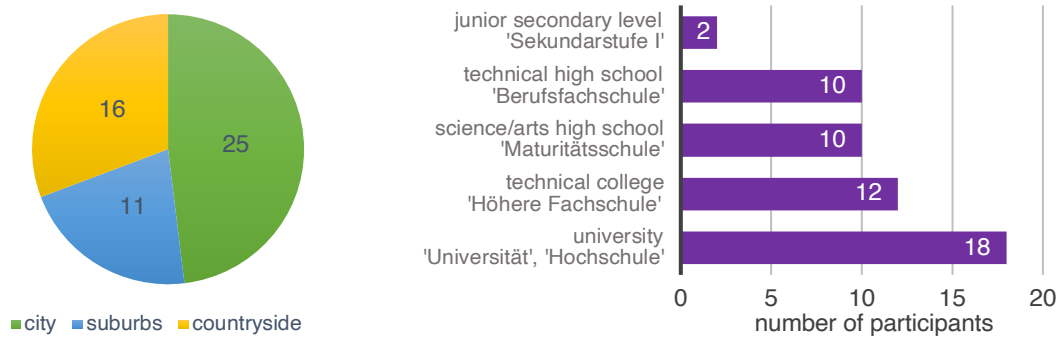

***Figure 3.*** *Subjects' place of residence (left) and highest completed education level (right).*

We favored group discussions over individual interviews for multiple reasons. They naturally allowed participants to confront differing opinions, helping to identify themes relevant to many. This stems from the belief that opinions and policies are products of social processes and negotiations, not isolated perspectives (Ferrara, 2018; Lamertz, 2002; Leonardi & Barley, 2010). We aimed to simulate this social exchange in our discussions (Bader & Rossi, 1998; Gaskarth, 2006; Leonardi & Barley, 2010; Puchta & Potter, 2004), hence groups included individuals from various backgrounds. We did not enforce agreement, allowing free expression and reference to others' opinions. Participants could extend on others' views or remain reserved. The field trial provided a shared scenario for participants and exposed them to the same actions of drones, facilitating deeper engagement with situational cues and specific episodes. This distinguishes our study from others focusing on drone acceptance using demographics or generic context. Group interviews also allowed two trials per day[3]. The shared experience aspect (as opposed, e.g., to discussions which would have been based on vignettes or individual experiences) was central to our study, as it is essential for opinion and policy formation (Ferrara, 2018).

---

[3] On average, 12 officers participated in each trial session, which generated major time and resource expenses. The command center was available for two days only. Thus, it was important to complete the field trial in two days.

We aimed to capture this effect in our approach. Overall, we sought to capture opinions emerging from a shared experience via social exchange, making group discussions the ideal method.

## 3.4. Data analysis

The recorded discussions were transcribed using the intelligent verbatim standard. The transcripts were analyzed according to the exploratory interpretative paradigm (Saldaña, 2009; Stebbins, 2001), i.e., the researchers aimed at establishing a shared interpretation of the data rather than objective correctness. They employed multiple measures to assure validity and reliability (Table 1). Exploratory research was acknowledged as a legitimate way to approach underspecified or complex phenomena that might emerge through the application of new technologies in the DSR, e.g., in contexts involving previously unexposed individuals (Briggs & Schwabe, 2011). The role of the researcher is to offer a potential explanation to the observed phenomena that can be used in subsequent DSR iterations.

Data was analyzed in two stages. Initially, a researcher identified key topics guided by a workshop held in August 2021. This workshop, involving four police representatives and four IS researchers, reviewed preliminary observations from the police's viewpoint. The workshop confirmed and specified the police's interests, focusing on understanding the perception of drone operations and potential bystander concerns that could hinder future operations or cooperation. The first author conducted bottom-up coding of bystanders discussions, identifying 522 relevant passages under eight different codes: police benefits (34 segments), bystander emotions (51), drone use adequacy (128), responsibility (75), recognizability (56), drone use value (15), transparency (48), and additional considerations (115). See the supplemental material for the codebook. The coding's accuracy was ensured through iterative checks by a second researcher, with any edge cases resolved through discussions.

***Table 1.*** *Steps taken by the research team to assure the reliability and validity of the research presented along the dimensions proposed by Merriam and Grenier (2019) – first six rows – and Lincoln and Guba (1985) – last three rows.*

| Strategy | Implementation |
| --- | --- |
| Member checks | It was not possible to conduct member checks with the study participants because of the necessary anonymization of the data. However, we consulted a preliminary set of results and observations with the police in a presentation and discussion. |
| Peer review/ examination | The obtained results were consulted with the researchers who participated in the field trials and data collection to ensure that the perceptions were consistent across the research team. Additionally, an external reviewer who is an expert in legitimacy and authority critically reviewed the paper prior to submission. |

| Researcher's position or reflexivity | The researchers carefully attended to this point and identified their main motivation. The authors are not funded by the police or any external party that might have an interest in promoting or discounting the use of drones. The funding for this project is provided by the national science agency (see Acknowledgement). The project is related to identifying the impact of digital agents and AI on political authority and legitimacy. The main motivation was to explore the application of drones in policing, particularly the impact of drones' use on the sense of safety in bystanders and the general acceptance of drones in law enforcement operations. The secondary motivation was a timely, informative publication on this topic. |
|---|---|
| Adequate engagement in data collection | Both researchers and the supporting research team were involved in the preparation and execution of the field trial, including contact with the police. They observed the participants during the trial sessions and moderated the subsequent discussions giving them a chance to refer to specific episodes during the discussions. |
| Maximum variation | As presented, the participants come from different backgrounds (education, age, residence). During the discussions, they were confronted with different opinions and could relate to them. |
| Rich, thick descriptions | The manuscript provides an in-depth exploration of the data collected and analyzed by the researchers, with excerpts from the data demonstrating the findings and their implications. This allows the reader to assess the conclusions drawn from the data and their implications. |
| Audit trail | The selection and analysis of the data were meticulously documented and controlled, thus ensuring that the process and decisions taken along the process are reproducible. The manuscript serves as a display of the analysis process and exemplary results. Specifically, many steps conducted by the first author were audited by the second author concerning their dependability, credibility, and confirmability. The audit trail was also extended by including police and other researchers in various workshops. |
| Dependability checks and accommodation | The methods employed were encouraging participants to openly communicate disconfirming opinions or alternative views. Negative evidence and conflicting views were collected and processed during coding and subsequent analysis. The Results are presented accordingly indicating what *themes* emerged and how often and showing variety of opinions concerning advantages of drones and their impact on safety. The adherence to those accommodation strategies were continuously checked throughout the iterative process of sensemaking oriented towards identification of contradictory cues. |
| Confirmability checks | The selection of methods and the design of the scenario to be used in the field trial were conducted in cooperation between the police and researchers of various backgrounds. The process was carefully documented and later reviewed for identification of possible gaps. Also, the analysis was subject to completeness checks by the second author, who was not involved in the early steps of coding and clustering, and in workshops involving researchers from authors' group who were involved in the field trial as assistants. |
| Credibility checks | Credibility checks were applied to control adherence with strategies such as member checks, variation, or peer review. Those check were discussed throughout the process, especially in the workshops with authors' research group and in the workshop with external IS reviewers. |

In the second stage, the authors, along with other researchers, participated in iterative sensemaking and data restructuring. This involved asynchronous exchanges via a shared repository and two interpretation workshops with the authors' research group including individuals who supported the field trial as assistants. This method allowed the authors to assess the completeness and authenticity of their interpretation, enriching the potential meanings of the selected content. This ensured the results were not solely based on the authors' individual perspectives but emerged from an intersubjective process. This sensemaking identified two meta-classes: attitudes and concerns. Within these, clusters and categories were identified, as detailed in Section 4. This required a new structure to be imposed on the previously coded data. The category "benefits for the police" was divided into multiple sub-categories (Table 2), and the "impact on bystanders emotions" was divided by the various emotions addressed in the statements. Safety, being the most prominent emotion in the data (30 out of 51 statements), was the main focus, with statements separated according to their valence (Table 3). Other emotions mentioned, such as disturbance, annoyance, and distraction, related to drone noise, were deemed irrelevant by participants for the scenario tested and were therefore not included in the results. Finally, codes related to the appropriateness of drone use, its value, recognizability, transparency, and other considerations were regrouped according to participants' descriptions of adequate drone use and how they differentiated it from inadequate use (Table 4, 336 coded segments). The meanings of the identified categories are described in the results, referencing the group discussion (A through J).

## 3.5.    Development of policy propositions

The primary aim of this research was to propose extensions to the existing policy. The authors used an iterative interpretation process based on the results outlined in Section 4. The police's interest in understanding the legitimacy of drone usage from the perspective of bystanders and witnesses prompted us to refer to the police legitimacy literature. This literature emphasizes the importance of safety feelings and justice perceptions in attributing legitimacy and fostering cooperation with the police (Bolger & Walters, 2019). We utilized the organizational justice framework, given its comprehensive nature and broad applicability, to structure our interpretations and formulate appropriate guidelines.

The refinement of the proposed guidelines was conducted through two workshops. The first workshop in February 2023 included a professor, three advanced doctoral students from the chair of political philosophy at the authors' University, an information systems doctoral researcher, and the authors. These participants were collaborators focusing on the legitimacy of drone use in safety-related contexts. The proposed guidelines were part of the project's deliverables, hence all participants had a vested interest in their development. One participant, an expert in norm theory, had previously assisted in developing official policies on property rights. Participants were given an early version of the guidelines and the reasoning behind them prior to the workshop. During the workshop, the results and proposed guidelines were presented, followed by a 45-minute discussion. Participants were asked to evaluate the guidelines based on the trial's results. They were given time to note their concerns for each guideline in a shared virtual space. These comments were later collected and discussed. The workshop was recorded for documentation. The philosophy experts suggested improvements to the normative style of the guidelines, indicating how they should be formulated for policy development. Questions were raised about the difference between responsibility and justice, the dimensions of justice, and individual guidelines. Despite these discussions, participants agreed with the structure and specific rules. The workshop concluded with minor adjustments to the formulation and scope of specific guidelines.

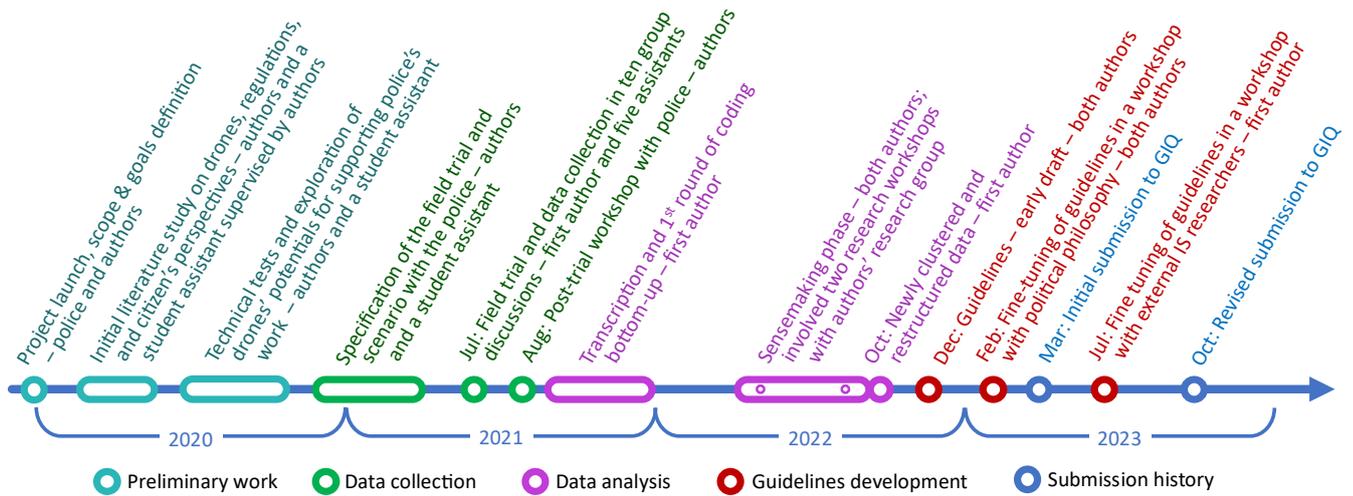

**Figure 4.** *Project timeline indicating activities, participants in those activities, and responsibilities.*

The second workshop, held in July 2023, involved nine information systems (IS) researchers from a European University specializing in digital transformation, cryptocurrency legislation, security, and energy policy. These researchers were external to the project and our University. Among them, two had substantial experience in fairness; five were postdoctoral researchers, and four were doctoral students. The workshop aimed to gather external perspectives on the proposed guidelines and assess its understandability and usefulness. It lasted an hour, featuring a presentation of the guidelines, followed by a 30-minute discussion. Participants critically evaluated the logic of the guidelines, commenting on the project's methodology and rationale. This discussion prompted reflection on the process and helped us contextualize our study within the Design Science Research (DSR) framework.

Overall, the workshops offered an early expert evaluation in accordance with the DSR practice (Hevner, 2007). They led to the improvement of the catalogue itself (first workshop) and the overall presentation of the study (second workshop). Figure 4 presents the timeline of the project from the initiation in early 2020 up to the revised submission made to this journal.

## 4. RESULTS

The results are based on the recorded, transcribed, and coded group discussions from the trial. These discussions effectively stimulated interaction among participants, who inspired each other and explored new perspectives under a researcher's guidance. The interview questions or statements referred to trial experiences, highlighting the participants' roles as witnesses and bystanders. Still, participants subtly assumed varying roles in the discussion, with some statements reflecting the viewpoint of a witness or bystander, while others were more ambiguous. Some statements reflected a broader citizen perspective, indicated participants empathizing with the police or perpetrator, or referred to hypothetical situations inspired by or contrasting with the trial experiences. Participants sometimes drew on their prior knowledge or experiences, particularly when adopting a citizen's perspective. We include such diverse statements in this section. We begin with participants' general attitudes and the benefits they attribute to drones, as well as their self-reflection on feelings triggered by the drones. We then discuss themes related to the legitimacy of drone use, which constitutes the majority of the results.

### 4.1. General Attitude

The analysis indicates that UAVs' use for policing is accepted. Only one person argued against it based on their principal view of drones as useless and dangerous. The remaining 51 supported the idea of using drones by the police mostly referring to the experiences from the trial or assumptions they made about what they saw in trial based on their prior knowledge. All of them see benefits for the police (cf. Table 2). They stress that the drone provides a better overview and more flexibility in conducting observations, leading to a faster response and more safety for the involved officers.

Subjects agree concerning the positive impact of drones on the police's work. Yet, they disagree about the impact of drones on the uninvolved bystanders or society. Some subjects assume that whatever is good for the police is good for the people: "[I see a drone as] a resource. If I know that the police use that, then I think, ah, is good, use that. It certainly helps" (G). However, most participants had differentiated thoughts about how they felt when drones were used and the implications of it.

***Table 2.*** *Identified advantages of drones for police.*

| Advantage | Exemplary Statements |
|---|---|
| Better overview (13 segments in 8 group discussions) | "You can cover a large area from high up with a drone. And through that, you can also already see someone who has fled the house and is a bit further away. Yes, so basically positive." (B) |
| Faster response<br>(3 segments in 2 discussions) | "And I think it's actually good that they can now see things from above and can go after the criminals more quickly and grab them." (D) |
| Higher flexibility<br>(4 segments in 3 discussions) | "In Germany, where I saw it, they collaborated: helicopter plus the drone. The drone could then get even closer because the helicopter cannot get down from a certain height. And from that point of view, I think that is really very helpful." (J) |
| More safety<br>(8 segments in 3 discussions) | "And (...) also positive, fewer people are endangered; also fewer policemen are endangered, so from that point of view I see it positive." (D) |
| Reliable records<br>(6 segments in 4 discussions) | "What I could also imagine in a police operation; if someone is arrested and then accuses the police of having been too brutal, then the film recordings could possibly help to clarify that." (E) |

## 4.2. Impact on Feeling of Safety

Asked about the impact of drones on their sense of safety as bystanders or witnesses, the participants provided various opinions (Table 3). Some appreciated having someone with an overview of a potentially dangerous situation to minimize potential risks. Those expressing ambivalence did not see how a drone could intervene if there was an immediate danger to their health or life, so they did not see the drone as relevant to their sense of safety. Finally, those feeling threatened by the drones most frequently cited the feeling of being watched, the lack of knowledge about the drone's task or origin, and, as mentioned, their principal attitude towards drones. Overall, the feeling of (un)safety was referred to the most compared to other emotions: 30 out of 51 statements on emotional impact of drones were about safety. The other 21 statements were referring to the feelings of disturbance, annoyance, curiosity, distraction, a mix of those, or unspecified emotions ("it felt a bit strange").

*Table 3. Impact of drones on subjects' sense of safety*

| Impact | Exemplary Statements |
|---|---|
| Positive<br>(11 coded segments in 7 group discussions) | "Now, if that was really true, a dangerous offender or so, I would have confidence in this whole thing, in the police effort, because they were supported by this surveillance from above. (...) As soon as I put myself in the situation, I thought, 'thank God, the one up there sees everything'." (J)<br>"If someone disappears between the houses, then it is hard to find them. So, with a drone, you can follow them quieter and faster than if one needs to run after them. So, a little bit more security, I think. (...) There is already a bit more safety." (E)<br>"It gave me a sense of security. I have the feeling that this is extremely supportive of the police, that they have an overview of where things are happening, and I think that gives you a sense of security." (B) |
| Neutral<br>(13 coded segments in 7 discussions) | "Not more security, there was not. So, the drones can't do anything. If the criminal, I don't know, has a bomb in their backpack or comes at me for some reason, the drones won't help." (E)<br>"Well, you see a lot of drones. It's not something special in that sense. But I didn't feel safe because the robber could have come and shot me anyway. So, then the drone probably doesn't do me much good either. But probably it has, I think, an influence on the robber, too, when he sees the drone. Then he knows he could be caught faster." (I)<br>"It is certainly important for the work of the police. But (...) just thinking about us standing there, at that moment, (...) it is not doing anything for us." (F) |

| Negative (6 coded segments in 4 discussions) | "Interviewer: What was the impact of the drone on your sense of safety? Subject: A negative one. I felt observed, and I think I didn't present myself or act the way I would be without monitoring. (...) As I said, I feel like I'm being watched by someone I don't know. That makes me uncomfortable." (F) "So, I actually also felt disturbed (...) because we're just not used to that yet. I think we are not yet used to the fact that the drones are for the good, because we are not yet far; it's not yet so normal; it's just still a bit different. And this perception of us is just already that a drone disturbs because you do not know what it does, what it is there for." (D) |
|---|---|

Generally, participants concur that drones can significantly enhance police work, making it more efficient and accountable through recordings. However, drones can also evoke negative or mixed emotions. All participants suggested ways to improve or limit drone usage to make it more acceptable to citizens, particularly bystanders or witnesses. We will now analyze these suggestions.

## 4.3.    Conditions, Constraints, and Considerations

The subjects developed various feelings towards the drones based on their participation in the field trial. In the discussions, they indicated manifold themes that need attention before the drones can be put into regular use by the police. They were frequently discussing under what circumstances the use of a drone would be adequate, appropriate, or reasonable ("angemessen," "sinnhaftig"). Their comments pertain to technology, its operation, and societal and regulatory aspects. Table 4 lists the most prevalent themes, along with exemplary passages from the group discussions.

***Table 4.*** *Criteria for adequate use of drones by the police.*

| Theme description | Exemplary Statements |
|---|---|
| **Technology** ||
| **Identifiability as a police drone** One of the drones in the field trial had a police-typical color and marking, which, however, might have been difficult to notice due to sunlight or altitude. Many participants made clear that it would have been necessary to see that the drone belonged to the police and was used for an actual mission, especially because they arrived earlier than the police officers. (35 coded segments in 10 group discussions) | "I also thought, for example, that it was not marked, that is, not somehow in the police colors or something, that it belonged to the police. It could have been a private person. That's why I didn't perceive it as safety. (...) That's why I wouldn't have felt that as security now because I can't assign it whether it's private or police." (D) "So, for me, it would still be important that this drone, that you can identify it, that it is from the police or so. But how that works, I have no idea. I'm rather "upset" by the drones, which are toys. That's annoying. On the other hand, if it has a function, then I think that's right." (A) "I was confused at first because I couldn't connect the drone with the police because they did were not simultaneous as I perceived it." (B) |
| **Reliability of technology** The participants discuss the reliability of the technology as a relevant aspect. They acknowledge that technological developments have been made, but they are afraid of mistakes it can make, be it while trying to identify the perpetrator to automatically track them or | "We have already gone far with the technology; it is very developed. And then I wonder (...) how far is this really well programmed or equipped? What happens if the program crashes? What can happen then if that were in the middle of a mission?" (J) "A: So, if [the drone] then follows, suddenly, someone wrong, because you simply have the same characteristics. You have the same height, the same stature, and maybe the same hair color, and then |

<table>
<tr>
<td>
landing without considering the by-standers. Even if they know that humans might also make mistakes, they consider a human mistake less problematic than one made by technology.

(11 segments in 5 discussions)
</td>
<td>
that can be problematic. And I just know, facial recognition and things like that, they're just not as high-tech as they probably need to be (...)
B: The policeman could make exactly the same mistake. He would also say, 'hey, look, you're the same'.
A: Yes, but I'd rather have a person make a mistake than technology.
Interviewer: Ah, O. K., why is that?
A: I don't know. I'm rather people-oriented than technology-oriented." (C)
</td>
</tr>
</table>

| Operation |
|---|

| | |
|---|---|
| **Identifiable target/function**<br>The inability to recognize the target and function of the drone also contributed to the participants' confusion. Concerning the function, they wished they understood what the drone was doing and how it was related to the police's mission. Concerning the target, they particularly stress the importance of knowing that they are not in the focus of a mission or the use of drones therein. They establish suggestions on how this could be achieved, e.g., via the use of cellular infrastructure.<br><br>(46 segments in 9 discussions) | "I would have liked someone to actually say, 'Hey, there's a drone. You guys are safe. It's not because of you; it's because of them.' And I think that even if a drone is used, I think it's important that the people who are being watched, which we inevitably were, are also informed. 'The drone is now there for that and not because of you'." (D)<br>"It was not exactly obvious to me what the function is. What it is doing. (...) Because it came before [the police officers], and we were then much later actually asked to move away. But, also, afterward, the function [was] not quite clear." (A)<br>"I think what would be useful would be a combination with other things, for example, that there is an info on the cell phone, 'There is something going on here right now.'" (H)<br>"Then it would perhaps still be important that one knows that. (...) you would just have to be informed. 'Hey, this is going on right now.' So you know a little bit. So that you don't just think, 'hey, they've been flying over the neighbor's house for two hours. What's going on?!' We might have to worry about that and not sit outside in the backyard anymore." (H) |
| **Use's duration and timing**<br>Some participants differentiate between permanent and punctual, timely limited use of drones. Accordingly, they argue against the use of drones for completing durable tasks of the police, including patrolling. They relate this aspect to their privacy rights but also to the preference for human contact with the police.<br><br>(34 segments in 8 discussions) | "I think the duration of the pursuit still plays quite a role. Because today, in this mission, it was maybe 15 minutes or so, or maybe 20. And if the person who is at a distance [the evader] is the only person that's being tracked, then it's really tied to a mission." (E)<br>"But I don't think they should fly around all the time instead of [police's] patrolling on foot. Surely – use from time to time, but not to replace the patrol on foot. (G)<br>"Well, I also think that if the police were to use it, such as for patrol officers who drive through the area from time to time, I think I would have more of a problem. But if this is really used selectively for the mission, then I think it's a good tool." (H) |
| **Safety of uninvolved persons**<br>The subjects point to the safety risks for the bystanders when considering the adequacy of drone use for policing. In this regard, they particularly refer to physical safety, for instance, which might be affected by a drone's falling down.<br><br>(26 segments in 8 discussions) | "I also think it has to be a balance of interests between what you want to achieve and the disadvantage for the other people who are affected (...) And if nobody is affected, then you can use it in a low-threshold way. If it takes place somewhere in a busy area, then the interests must be weighted higher in order to be able to justify it." (B)<br>"Safety must be guaranteed. If it crashes, it must somehow have an emergency program that it does not fly into the mass. That would certainly still have to be guaranteed." (E) |
| **Efforts and costs**<br>Most participants see a relation between the advantages of a drone's use (cf. Table 2) and the effort or costs related to use. They call for a careful balance between those aspects or directly express a preference for other, more traditional means of policing.<br><br>(9 segments in 4 discussions) | "I think the effort must simply be in proportion to the benefit. So, if you can do something just as well with other means, then the benefit just has to be high. Not that you're always flying around with this high-tech thing when you could do it differently." (B)<br>"I don't know what something like this costs and what else could be done with the money. So, let's assume there are fixed expenses. I'm not in a position to judge that. But I would hope that the police would look at it carefully and say to themselves, the benefits that we get from this outweigh the costs. Then I have confidence in that." (D) |

| Society | |
|---|---|
| **Societal benefit**<br>In deciding whether a drone's use is adequate, many people consider the broader benefit it generates. If the drone creates an added value, compared to the status quo, they see it as the right mean to be used. Some see convicting the offender as the highest value, even compared to their own or the bystander's rights. Those subjects differ from most of the subjects whose opinion is discussed below.<br>(38 segments in 9 discussions) | "If it allows the police to catch the perpetrator faster. Or if they have the overview on a huge property or whatever, and they would need the help to catch [the offender] faster. Or if it's really a serious criminal or something. To make sure that he can't escape or walk around free, I think so." (I)<br>"They should take the means, which is more efficient and leads better to the goal, I would say now. For me, it doesn't matter if they are over my garden or in my garden, but simply if they have caught the person they were looking for. That is the fairness for me that they then catch the one who robbed the jeweler." (C)<br>"I don't think the impact is that serious for non-participants; I think it's actually also justifiable if it helps to clarify the situation or if it can make a contribution to the clarification of a crime, regardless of what kind of crime it is." (A) |
| **Transparency and information**<br>The subjects see it as necessary to inform society about the use of drones by the police. They claim this might enhance the acceptance of drones and reduce the confusion if drones arrive at a crime scene. Some link this aspect with the public debate about the rules for the use of drones in policing.<br>(33 segments in 10 discussions) | "But if I know in the future, for example, that the police have drones, today, I don't think you have any. Then I also know what it is. 'Ah, that is probably police, then it would not worry me'. So, you would certainly have to inform people." (G)<br>"There is definitely a lack of information or communication. It comes down to the same thing, (…) that you know where this drone comes from, I think that's missing. I also [think] that it would be good to have an information campaign or something. Because for us, it was apparently new. In the future, if it becomes a reality then, then it would be good to know so that you don't have like, 'oh, where did that drone come from, or is that a private drone or so'." (D) |
| **Gravity of the offense**<br>The subjects refer to the balance between the impact of the offense and the use of technology. They consider using drones or other technical tools for the prosecution of "small" offenses (speeding, parking, consuming soft drugs) unproportional. Most of the participants also do not see prevention as a legitimate use scenario unless there is a major danger to a large group of people, such as during a parade. They list such offenses as criminal assault, robbery, or riots at non-peaceful demonstrations as appropriate scenarios for the use of drones.<br>(83 segments in 10 discussions) | "If it was a petty offense, a minor theft or something, I don't think it's necessary. To drive up such a huge contingent. And with something bigger, with a bigger crime, you have to intervene properly." (B)<br>"I think [using drones is OK] when something is really going on. Not proactively just fly around a bit and see what's happening, but when there's really a gathering of people or a robbery or something like that, that's when the drone comes into play. But not prophylactically." (E)<br>"So, up to a certain dose, you should also be able to [use a drone against] a drug dealer. And, if it is obvious, then the police will catch them. But if they come with drones (…) when one [has] two joints, then that would be too much for me." (I)<br>"I believe that there is a need for clear basic legal provisions that can be used to prevent or investigate serious crimes. We're not talking about a purse-snatcher who just quickly steals a wallet. We are talking about serious or massive violent crimes or anything else that would have a major impact on a population group or society in this area." (J) |
| Regulation | |
| **Regulation of use**<br>The subjects suggest that rules need to be in place to balance the interests of various stakeholders. Many see the use of drones by the police as a way to narrow down the freedoms of individuals suggesting the need for a referendum. They are, specifically, afraid of the misuse of drones for all-encompassing surveillance or unjust use of those additional means. | "Then we must make an initiative in Switzerland concerning the use of drones." (I)<br>"So everywhere where I am curtailed in my freedoms, in my fundamental rights, I think a drone has no place. Unless you can clearly declare it to be used to thwart serious crimes or for crowd control at a street parade, where you know that it's all about getting people out of their homes." (J)<br>"It might be that dark-skinned people, young people are more likely to be observed (…) one talks about profiling and such things. Then, I think, independent of the technical progress, but these are still those people who feel this, and there I have not only ethical but also legal |

| (20 segments in 7 discussions) | questions. (...) I think it should be clearly defined and somehow anchored, perhaps by law. We don't want this to get out of hand like all the technology nowadays and with artificial intelligence." (I) |
|---|---|
| **Privacy and data protection**<br>Many subjects connect the use of drones with surveillance and do not like the idea of being observed or tracked. However, this position does not lead to a fundamental refusal of drones in policing but rather to calls fa or a balance between privacy rights and drones' advantages. The subjects differentiate between justified and unjustified mainly along the lines of crime's gravity), involved and uninvolved individuals, and extreme situations and daily life.<br>(21 segments in 8 discussions) | "When I'm in everyday life, then I don't want to be permanently recorded somewhere or somehow... Because I think it's nobody's business where or what I do. If I don't do anything wrong, or I don't do anything criminally relevant." (J)<br>"You just feel observed with a drone, no matter when, where. We also feel observed now, but [it is] justified. There is something important to observe. But parking ticket penalty or so, you feel disproportionately observed and controlled." (H)<br>"So, I think we're still not so used to this kind of thing going on with drones. For me, a drone is still surveillance; it monitors you, so to speak. Big brother is watching you. That was my first thought." (D)<br>"If I'm recorded somewhere where maybe I'm just passing by or something, and I'm not involved, I would think that's unfair too, that I'm being observed there." (J) |
| **Right to be forgotten**<br>The subjects attend not only to the moment of data collection but also to what might happen with tracking data or video recordings afterward. They expect that the police would delete the collected data when it is not relevant to the original offense anymore, such that it is not being used for another purpose.<br>(6 segments in 3 discussions) | "I think it should be used a lot. But this data should be deleted after a certain time. So, if we now film a festival, street parade. Then it should be deleted after a certain time. Not that you can still trace what happened years later. But you should use it; you should need the technical means. And sometimes you see, just this drone, that is filmed, and then you can perhaps in meticulous detail, let the film run slowly, you see actions in retrospect. But that should be deleted again sometime." (D)<br>"If I'm on a picture somewhere, no one is interested because I wasn't exactly in danger. Then I don't care if it will be deleted again." (J) |

They provide numerous suggestions on how to increase the likelihood of bystanders and witnesses attributing legitimacy to drones and the police. They also discuss ways to minimize the negative impact on the citizens. Notably, despite varying views on potential impacts, the suggestions were consistent across the data set. Nine out of twelve themes were discussed in seven or more groups, indicating their significance to the participants. There were no clear patterns linking participants' individual characteristics to their statements. Opinions were formed through individual interactions. Each group discussion had its own unique dynamics. This adds to the qualitative depth of the collected material and aids in uncovering nuanced meanings identified by some groups (such as the difference between data protection and the right to be forgotten, or various aspects of identifiability). However, the data is not suitable for quantification beyond simple theme frequency.

## 5. TOWARDS A POLICY FOR JUST DRONE USAGE

The findings show, participants accept drones as an aid for the police – similar to results obtained in the past (Klauser, 2021b). Yet, acceptance comes with limitations. Study participants say drones' use will improve the state's power which may generate conflicts with individual rights. They expect that state-wide regulations and police routines will curtail the power that comes with new technology. They call for rights that strengthen the position of an individual, thereby supporting prior recommendations (Bentley, 2018; Fox, 2017, 2022). In general, citizens do not want the state to employ modern technologies without considering the affected rights (Heen et al., 2018; Sabino et al., 2022; Sakiyama, 2017) and our data supports this insight. The following presents how existing policies and regulations can be extended to meet expectations bystanders expressed in our trial. Later, we discuss the implications.

In alignment with the goal of the study, i.e., development of propositions to supplement existing policy, we used the analyzed data to develop a catalogue of guidelines to inspire adequate policy. Frequently, a single guideline sources at several observations presented in Tables 2, 3, and 4. We indicate those references through *underlined italics*.

When explaining their rationale, the subjects refer to appropriate, reasonable, or fair drone use. Organizational justice concerns interaction, procedure, and distribution (Greenberg, 1986). The participants' statements attend to all dimensions of justice and generate wide-ranging implications. In the following, we discuss those aspects and list them in Table 5 while proposing *catalogue of guidelines for just usage of drones (*short: *just drones guidelines)* to acknowledge bystanders' needs. The guidelines are recommendations for further development of the existing policy. We intend to offer those guidelines as indications for how policymakers and police can ensure a that drones will be accepted in the eye of witnesses and bystanders. We claim that, by employing those guidelines, the police will be able to ensure that those individuals will accept the use of drones as a legitimate way to exercise power.

Procedural justice assures that rules and processes are consistent, comprehensible, and traceable (Colquitt et al., 2005; Schaap & Saarikkomäki, 2022; Tyler, 2021). The subjects demand that the police use of drones should be subject to comprehensive legal frameworks, reaffirming calls for such laws (Fox, 2022; Heen et al., 2018; PytlikZillig et al., 2018; Sakiyama et al., 2017; Smith, 2015; West et al., 2019). Generally, the participants request a clear *regulation of use* to balance various stakeholders' interests. They demand rules for *privacy and data protection*, as well as *right to be forgotten*. They also refer to regulations to assure adequate *gravity of the offense* or *duration and timing* for the use of the drone. Finally, they are unsure about the *reliability of the technology* and expect clarification of edge cases responsibility. To address this theme, we offer *regulation guideline* to address procedural justice.

Interactional justice has two aspects: interpersonal and informational justice (Bies, 2001; Colquitt et al., 2013; Colquitt & Rodell, 2011; Dolata et al., 2022). Regarding the use of drones, interpersonal justice ensures that citizens are treated with respect and dignity, whether by the drone or by the police equipped with a drone. The subjects demand interpersonal justice when claiming *safety for uninvolved persons*, *privacy and data protection*, and *the right to be forgotten*. In all those cases, they explain how their rights as individuals who are not involved in the crime might be reduced by using drones. We argue that the rights of those who, without fault, happen to be at a crime scene must be secured, given that they might already be traumatized by unwillingly participating in the event. Accordingly, we propose *considerateness guideline* and *confidentiality guideline*.

Informational justice ensures citizens are given necessary information on what has been undertaken to assure justice (Colquitt et al., 2013; Colquitt & Rodell, 2011; Robert et al., 2020). Subjects want a drone that is *identifiable as a police drone* because they trust police rather than private owners concerning compliance with the law – this confirms previous research (Ahrendt, 2020; Aydin, 2019; Boucher, 2016; Eißfeldt et al., 2020; Lin Tan et al., 2021). Also, they expect that they – as witnesses or bystanders – should be able to clearly *identify drone's target and function*. The participants request *transparency and information* about the police using drones to be prepared for such an intervention and to ensure their rights are respected. We derive *recognizability guideline* and *transparency guideline*.

Distributive justice assures that necessary resources are allocated adequately or equally (Colquitt et al., 2005, 2013). Similarly, positive or negative implications of drone use should be fairly distributed. The participants indicate that resources used by the police must be proportional to the benefits and risks. Police should consider a complete set of societal needs and not just safety. The data indicates that *the gravity of the offense*, *use's duration and timing*, *social benefit*, as well as *efforts and costs* play relevant role in participant's assessment of balance between advantages and disadvantages concerning drones' use impacting the attributions of legitimacy. With the gravity of the offense and use's duration and timing, the participants demand a balance between their rights (e.g., privacy, freedom from being suspected of a crime) and the police's right to control the law adherence. With social benefits, efforts, and costs, the subjects demand that the resources used for policing are proportional to the positive outcomes to be achieved. In response, we propose *balance guideline* and *sufficient benefit guideline*.

***Table 5.*** *Proposed* just drone rules *to acknowledge bystanders' needs during police operations involving the use of drones.*

| Guideline | Specification |
|---|---|
| **Procedural Justice** ||
| **Regulation guideline**<br>Clear procedures, processes, and practices need to be codified, implemented, and regularly assessed. The regulations should attend to all aspects of justice, i.e., assure the implementation of other principles listed below. They need to be applicable across situations and over time. The legitimacy of the regulation needs to be assured. | As shown, there is a need to balance multiple interests when using drones for policing. This cannot be done without an appropriate legislative process considering those interests. The regulation needs to provide guidance to the police, uninvolved individuals, and further persons who, e.g., might work with the material collected by a drone. Standards must be in place to guide the development of drones and ensure the technology is reliable enough. Given that drones rely on probabilistic processing (e.g., visual computing to predict optimal paths), the risks that might occur due to technical imperfection need to be weighted accordingly. |
| **Interactional Justice: Interpersonal Justice** ||
| **Considerateness guideline**<br>The drone must be aware of, consider, and anticipate the presence, behavior, and potential needs of uninvolved persons at the crime scene during the operation. It should reduce the risks for them through its own actions. | The drone must be able to perceive the environment and interact with it in a way that reduces the risks for uninvolved individuals. It includes making predictions about its own future states, about the future states of the environment, and about its impact on the environment. For instance, indicating to the bystanders that they are not of interest as criminal subjects might reduce their nervosity and, therefore, the risk of unexpected behaviors. |
| **Confidentiality guideline**<br>The drone should, if possible, only collect and record data that is necessary for its working, the criminal act, and the security of the involved. The recordings should be used for processing only the crime for which the drone was employed. The data should be deleted as soon as possible. | One of the drone's advantages is its ability to collect multimodal data of high richness with added value for police. The dignity of and respect for the uninvolved individuals demands that the collected data will not be used to their disadvantage, be it directly during the operation or afterward. This can be achieved by implementing a limitation in what data is collected, e.g., observing and recording the offender's fleeing rather than the whole scene. Additionally, rules and automatisms must be in place to process and delete the data. |
| **Interactional Justice: Informational Justice** ||
| **Recognizability guideline**<br>It should be clear that a drone belongs to police or acts on its behalf. Further data about the drone (piloted vs. autonomous, information collected, etc.) must be provided to the bystanders unless it collides with the interests of security and safety of the police or society. | Recognizability of the drone as belonging to the police can be achieved through reserved light signals visible from far away, color, audio signals, specific movements, and mobile broadcasting messages. They need to be standardized, legally protected, and known to society. This is particularly important if drones arrive at the crime scene before the police to provide a sense of control and reduce confusion about the adequacy of drone use. Various signals can be used to indicate whether the drone acts in an emergency or conducts other tasks. |
| **Transparency guideline**<br>Publicity should be broadly and permanently informed about the use of drones, including information about their general abilities, potential use cases, relevant rules, and adequate behavior if someone spots a drone. This includes information about how citizens' rights are protected in drones operation. | Society has not yet established a broadly accepted code of conduct for interaction with drones. This enhances the risk of misinterpretations or even leads to hostility towards the police or drones if society is not convinced that the application of drones is legitimized and controlled. This information is central to establishing informational justice. This can be achieved by a broad information campaign, meetings with the citizens, demonstrations, or public debate. Politics and public agencies need to initiate a public debate and consider its outcomes for regulatory and informational tasks. |
| **Distributive Justice** ||
| Balance guideline | This guideline attends to the right of all persons to present and put forward their individual or group interests. This right needs to be distributed impartially. The police are allowed to prioritize safety and |
| The police need to consider the balance between an offense's gravity, the uninvolved individuals' rights, and the goals related to security and safety. This assessment might be guided by general regulations but requires case-to-case assessment. This implies that the responsibility for this assessment is clearly defined. | security over other interests if necessary. This also affects the decision of whether drones should be used at all, for which offense, and for how long, i.e., at which moment of the operation the reason of safety or security weighs less than other interests. Police are subject to similar rules in other areas (e.g., in what situations they are allowed to overcome traffic laws). The use of drones requires new procedures with similar characteristics. |
| **Sufficient benefit guideline**<br>The use of drones can be generally allowed if their benefits are confirmed by evidence and weighed against needed resources. This guideline refers to the general question of what conditions should be met that legitimize the provision of drones to the police. | It is necessary to understand what benefits (can) emerge from using drones by police. Those benefits need to be assessed against their societal relevance and efforts to achieve those benefits. This involves sincere and systematic study and control of drones' impact on law enforcement, e.g., in a field trial setting. If drones are to be rolled out to the police, their efficacy needs to be subject to independent control. Information about the efficacy of technological interventions is important to prevent unnecessary investments. |

# 6. DISCUSSION AND IMPLICATIONS

The findings and suggested guidelines highlight key issues regarding drone usage in law enforcement. Unlike broader studies, they also implies specific solutions or ways to identify them, and clarify public reservations and fears. We will present these insights, outlining their contribution to the literature.

## 6.1. Selecting adequate application scenarios

Previous literature identified many different applications of drones in policing. We selected one with good acceptance but not from the top of preference lists (Eißfeldt et al., 2020; Heen et al., 2018; Lin Tan et al., 2021; Sabino et al., 2022; Sakiyama, 2017; Saulnier & Thompson, 2016). Our research confirms that people find this scenario adequate and severe enough for drones to be used. They acknowledge that employing drones during a bank robbery and subsequent pursuit generates advantages not only for the police but also for society or bystanders, significantly impacting their attitude. In the existing research, the advantages of drone use are predominantly studied or presented from the perspective of the police (Anania et al., 2019; Klauser, 2021b; Sabino et al., 2022). Frequently, studies establish the advantages through argumentation rather than empirically (Aydin, 2019; Fox, 2019; Liu et al., 2022). We show that people were aware of the advantages drones offer to various stakeholders, such as additional *safety to the police*, and *reliable records,* even if those aspects did not play an explicit role in the scenario which was rather making use of *better overview of the situation*, *faster response*, or *higher flexibility*. Compared to previous literature (Klauser, 2021a), they identify reduced response time and reliable records as potential advantages. Swiss police, as a user of drones, can be assured of support from the public if drones are used for suspect tracking after a robbery.

Subjects also reference less suitable scenarios such as traffic control, routine patrols, or minor offenses like drug use. This is consistent with prior research suggesting a preference for reactive drone use over proactive policing (Heen et al., 2018; Sakiyama, 2017; Saulnier & Thompson, 2016). Trial feedback indicates how these attitudes might form. We hypothesize that citizens employ a legitimacy calculus, assessing the safety risk's implications and the drone's potential benefits to police, bystanders, or society. While individual aspects are found in previous acceptance studies, our data provides insight into the process linking these assessments (Gómez et al., 2015; Kähler et al., 2022; Legere, 2019; Lidynia et al., 2017; Tepylo et al., 2023).

Participants in assessments consider liberal fundamental rights, such as freedom of speech and protest, and the frequency of an offense, favoring exceptional crimes over routine business offenses. This could explain why drones are preferred for search and rescue operations over traffic monitoring (Çetin et al., 2022; Liu et al., 2022; Sabino et al., 2022). Traffic violations are common and usually have minor consequences, not directly impacting society. Conversely, search and rescue missions are rare but can greatly affect the health or life of the individual and their family. Therefore, police officers should assess the appropriateness of drone use, as they do when deciding to use physical force. Although drones don't directly impact citizens' physical or mental health, they can cause distress or fear and infringe on privacy rights. This suggests that the discourse on balancing privacy and drone use is moving in the right direction (Bentley, 2018; Fox, 2017, 2022), but it should also consider other relevant factors. The proposed calculus might be more complex, with situational aspects like the behavior of others, raised opinions, provided information, or how drone use rules were established (e.g., public vote) potentially influencing the assessment of drone legitimacy. Further research is needed to understand how people form opinions about police use of technology.

In the bank robbery scenario, subjects outlined restrictions and proposed drone functionalities. Unlike general concerns in prior literature (Çetin et al., 2022; Sabino et al., 2022), subjects didn't mention job loss risk for police officers or the drone being stolen, hacked, or misused. Noise or visual pollution was also not a primary concern. However, participants emphasized the need for transparency and information, both generally and during operations. They proposed making the drone identifiable as police, indicating its function and operation purpose. Existing regulations (EU, 2015, 2019a, 2019b; Fox, 2022) address identifiability, focusing on technological means like registration or chips, but these may not suffice for ground-level identification. Addressing this could minimize potential refusal of police drones, non-cooperative behavior from witnesses or bystanders, and trauma from unexpected, dangerous situations. Given the infrequent and brief interactions citizens have with police drones, as suggested in prior research (Heen et al., 2018; Sakiyama, 2017; Sakiyama et al., 2017), making them recognizable warrants attention from designers and users.

Specific to the use of drones for policing, the existing literature emphasized reservations about privacy, bias against marginalized populations, and surveillance (Bentley, 2018; Liu et al., 2022; Nelson et al., 2019; Sakiyama, 2017; Sakiyama et al., 2017). Our data confirm that those aspects concern the citizens. Concerning privacy, the participants suggest addressing them, e.g., by supporting the *right to be forgotten* or by regulations assuring that the material will be used only to prosecute the crime for which the drone was initially deployed. Some state-specific US regulations address this by limiting the value of those recordings as legal evidence (Bentley, 2018). However, different states have different regulations, which is at odds with the citizens', police officers', and researchers' wish for consistent, contemporary, and applicable legal rules.

## 6.2. Governing and legitimizing the use of drones

This study presents a fresh view on the legitimacy of police drone use. It suggests justice as the basis for drone use legitimacy. We contend that drones should only be used in a manner that respects individuals' rights, ensures impartiality, and assigns agency for proper drone use (Miller, 2003, 2021). This extends beyond merely enhancing police performance in crime control (Liu et al., 2022; Packer, 1964; Stampa et al., 2021). We propose that justice, particularly towards bystanders and witnesses, underpins the legitimacy of drone use, adding to the broader discourse on legitimacy sources (Adams, 2018; Peter, 2017). This paper primarily focuses on perceived justice and its impact on legitimacy, defined as a belief or attitude towards law enforcement agencies (Mommsen, 1992; Peter, 2017; Weber, 1964).

The data indicates the need for a balanced and clever regulation of drone use. The interpretation of the concerns in terms of justice yields a framework for the legislation. Yet, it also shows there is room for interpretation. The definitions of organizational justice (Greenberg, 1986) and procedural justice (Tyler, 2021) show that justice does not emerge only from the regulation but also its implementation in a specific case. Aspects such as respect, dignity, or transparency rely on punctual interactions with the police and specific officers. Those interactions were identified as essential for justice assessments, legitimacy, willingness to obey the law, and cooperation (Fielder & Murphy, 2022; Paternoster et al., 1997; Tankebe, 2013; Tyler, 2021; Wolfe et al., 2016).  Interaction with a non-human representative like a drone might impact the perception of the police's justice[4]. It is hard (if not impossible) to regulate interaction. We suggest two complementary approaches to ensure that regulations concerning drones and their deployment are just and this deployment is seen as just by the citizens.

We assert that police officers need training on effective drone usage to enhance law enforcement quality. This focus is on effectiveness and quality over efficiency, contrasting with the prevalent discourse on policing, which often prioritizes efficiency (Fox, 2018; Liu et al., 2022; Stampa et al., 2021). Operation commanders, drone operators, and decision-makers must understand the balance required when deploying drones. Intense, practical training will enable them to make relevant assessments swiftly and reliably. Intervention troops are well-trained in quick situation assessment based on incoming cues. However, operators of autonomous or remotely controlled drones may lack this experience. Also, heuristics developed for intervention troops may not suit drone operators. Drone operators and commanders have access to more information than on-site police troops. This access to more data cues increases the complexity of the situation and necessitates many morally relevant decisions.

Intervention troops learn to deal with complexities that result from their legal power and weapons through training. Training drone operators or pilots requires a similar approach beyond the technical abilities currently required by law (Bentley, 2018; EU, 2019a, 2019b; Fox, 2019). It is necessary to introduce dedicated training and certification for police officers working with drones. This implies that the police's drone operations are subject to a dedicated regulatory framework. Such a framework might relax some of the currently existing airspace safety regulations, e.g., the rule requiring a direct line of sight between the pilot and the drone, which might be inappropriate if the pilot would risk their life and impose more adequate regulations and certifications along the lines proposed in this article.

---

[4] In an institutional context, humans tend to transfer attributes of technology on the organization (Dolata et al., 2019). Interactions with technological artifacts might thus be crucial for justice perceptions of an organization.

We advocate for enhanced efforts to address bystanders' concerns about drone recognizability and identifiability. This can be achieved by designing technological solutions, such as visual or acoustic signals, colors, and flight patterns. For example, certain colors could be reserved for police drones, similar to police car designs. These features could help identify the drone's affiliation and purpose. Collaboration between human-drone interaction designers and legislators is necessary to establish a simple, recognizable human-drone communication language. Society has already established legally binding communication rules in various contexts, some of which are lawfully protected, while others are unofficial standards. Similar signs could be developed for drones. Appropriate technical solutions could ensure the right to be forgotten or limit data access to designated officers. Regulators must enforce minimal technical standards for policing drones to ensure bystander safety. Ultimately, drone producers and designers can significantly contribute to legitimacy of police drones use.

We claim that the governance of drones needs to embrace the complexity that emerges through using drones in police operations. Technology, particularly one that uses machine learning or autonomous components, significantly enhances the complexity of any situation (Dolata et al., 2022). This complexity grows even further if we add concerns about society and the dynamic sociotechnical environment. We call for researchers to embrace this complexity instead of trying to multiplicate superficial and static presentations of the acceptance (Eißfeldt et al., 2020), which we claim is a dynamic phenomenon driven by public discourse, personal experience, and interaction.

## 6.3.    Implications and directions for future research

This study holds significance for regulatory, social, and technical research. Legal and criminal justice research discusses drone-related regulation (Bentley, 2018; Fox, 2017, 2018, 2022; J. R. Nelson et al., 2019; Shenoy & Tyagi, 2022). We advocate for the development of effective regulations, acknowledging existing efforts (Fox, 2017; Klauser, 2021b; Shenoy & Tyagi, 2022). However, many proposals remain vague and challenging to operationalize (Smith, 2015). Potential questions involve: Is a special status of police drones in the legal framework necessary, and how can it be implemented? Are dedicated control routines necessary, and what would be their goal?

Social research and the humanities are called for a more profound and careful approach towards drones' acceptance and the factors contributing to it. The value of broad surveys is enormous to learning about public opinion and provides a basis for industry and governance to pursue (or not) use of drones(Çetin et al., 2022; Del-Real & Díaz-Fernández, 2021; Eißfeldt et al., 2020; Lin Tan et al., 2021; Sabino et al., 2022). However, it is limited when exploring social and psychological processes of forming the notions. We claim that acceptance of drones is subject to various drivers, including public discourse, specific events, and emotional responses – this aligns with previous research on the proliferation of novel technologies (Bradford et al., 2020; Dolata & Schwabe, 2023a; Mora et al., 2021; Rossler, 2019). We encourage the researchers to engage in research involving realistic and real-use scenarios, given that demonstrations and direct interaction with a drone can fundamentally change a person's perspective (Chang et al., 2017; Kähler et al., 2022; Miron et al., 2023; Oltvoort et al., 2019; PytlikZillig et al., 2018). We call for social science researchers to research what impacts the change of attitude to drones over time and what events might have particularly detrimental effects.

Finally, technology design and deployment research, including information systems and digital government research, should address the technical and socio-technical challenges. There is a need to make drones and their use socially acceptable while keeping the advantages for the police. This might involve designing new interventions, improving technologies, and proposing new interaction modes between drones and humans. Research needs to embrace drone policing as an important and future-oriented topic. Information systems and digital government communities' past interest in designing and understanding the relationship between law enforcement and technology might be a valuable asset here (Dekker et al., 2020; T. L. Johnson et al., 2022; Sakiyama et al., 2017; Sarkar, 2021; van der Giessen & Bayerl, 2022; Williams et al., 2018). We derive the following research opportunities from the current study: What processes are needed to manage this complexity for the citizens, police, and drone pilots or operators? What design aspects of a drone make it appear legitimate to the public?

Overall, using drones for policing creates a research area with tensions and conflicting interests. Balancing them can be achieved through the design of technology and new processes, effective communication with the public, and effective regulation beyond airspace safety issues. The challenges can be addressed only through a multidisciplinary and considerate approach. This study sets the stage for this collaboration by providing insights into citizens' concerns and potential extensions for existing policies and by bringing together relevant discourses from various disciplines.

## 7. LIMITATIONS AND CONCLUSION

The study isn't without limitations. A genuine field trial, while lacking control and limited in internal validity due to spontaneous events that can shift participants' emotions and attention, wasn't feasible in a public area due to existing laws. Instead, subjects visited an emergency services training area, potentially influencing results. Group interviews, where opinions surfaced during discussions, have their pros and cons, but they also subject identified patterns to social influence. These limitations were considered during our data analysis and interpretation.

To address the limitations of this study, several steps should be taken in future research. Firstly, the suggested catalogue of guidelines should be practically tested, moving beyond the expert review approach. A field trial implementing most of the guidelines could indicate their impact on the legitimacy of police drone use in specific operations. Secondly, triangulation could be achieved by testing the guidelines in various scenarios with different populations using a wider set of methods. This would provide a more comprehensive view and observe actual cooperation behaviors. Thirdly, the findings and guidelines could be validated in survey studies to test their completeness and validity against a broader population, potentially beyond the Swiss context. We encourage researchers specializing in large-scale surveys to continue this work. Lastly, we urge the community to consider further guidance as indicated in section 6.3, which suggests potential areas for new research directions. We believe that drones will be an important topic in practice and science, and digital government and IS researchers are ideally placed to explore this topic and inform policy development.

Despite the limitations, the study has implications for practitioners and researchers. Swiss police obtain an analysis of reactions to drone use in a realistic setting. They learn how citizens might react if the drone is used in real. Policymakers obtain a suggestion for a framework based on justice and legitimacy derived from empirical evidence. Finally, researchers in the field of e-government and information systems receive an insight into an emerging and fascinating field of research with much potential for socially important contribution. We are confident that the use of drones in policing will improve over time and that it deserves intensified efforts from all affected stakeholders.

In conclusion, this study offers several insights. It highlights justice as a key factor in legitimizing the use of technology by the police, bridging the gap between technology usage, police legitimacy, and perceived justice. We also identify factors pertinent to witnesses and bystanders in evaluating police drone use, noting that previous research overlooked elements like timing and duration or recognizability and identifiability. Additionally, we offer a guideline catalogue based on organizational justice dimensions, derived from our findings. This catalogue suggests new considerations for policy development beyond privacy and safety. We encourage further exploration of practical ways to implement these guidelines. Overall, this study contributes to digital government and IS research.

## ACKNOWLEDGMENT

We thank the following individuals for their support during the collection, analysis, and interpretation of the data: DK, YW, AB, and FC. Most prominently, we thank DW for his engagement during the field trial and initial data analysis, which he conducted as part of his master's thesis. We also want to thank the members of the police department in YZ for the effective collaboration throughout the project. The study was supported by YXZ under grant 123.

## SUPPLEMENTAL MATERIAL

### Codebook

The codebook emerged in the bottom-up manner in the initial stage of data analysis. To some degree, it was inspired by a workshop with the police and researchers who acted as assistants in the field trial. Importantly, the codes were re-organized and split in the second analysis step, resulting in the clusters presented in Results.

| Code (#occurrences) | Explanation | Example |
|---|---|---|
| police benefits (34 segments) | Identifies statements in which participants discuss what advantages for the police emerge from the use of drones in law enforcement operations. The code was later split into several subcodes as presented in Table 2, Section 4.1. | "You can cover a large area from high up with a drone. And through that, you can also al-ready see someone who has fled the house and is a bit further away. Yes, so basically positive." (B) |
| bystander emotions (51) | Statements expressing how the drones or drones' actions, appearance, or behaviour impacted participants' feelings during the police operation. The code was later split by emotions and the feeling of safety was split into multiple subcodes, cf. Table 3, Section 4.2. | "Interviewer: What was the impact of the drone on your sense of safety? Subject: A negative one. I felt observed, and I think I didn't present myself or act the way I would be without monitoring. (...) As I said, I feel like I'm being watched by someone I don't know. That makes me uncomfortable." (F) |
| drone use adequacy (128) | Statements expressing in which scenarios and under which conditions people consider the employment of drones by the police adequate if the subject explicitly refers to notions of *adequacy* or if it inspired directly from the question about adequate or *justified* use of drones. This code was split in the second stage of analysis and contributed to most findings in Table 4. | "So, I think it is *justified* when there is a certain event. And it's also justified when there are large gatherings of people. I'm talking about Street Parade, for example, or just some big folk festival and so on. Then I think it should also be used to really get an overview." (J) |
| responsibility (75) | Statements indicating who is responsible for the drones' good- or wrongdoing and why. This code was excluded from further processing for this publication, as it would add an additional layer of theoretical complexity. Further the answers abstracted from the experiences from the field trial and sounded indecisive. | "The person who programs the drone or the company that produces it may also have some responsibility, and then, the drone is controlled by a person, depending on the mission. (...) If it is autonomous, then it is the company or the enterprise or the police that uses it. (...) I think you have to weigh that up depending on the situation." (D) |
| recognizability (56) | Statements referring to participants' drone's recognizability as a police tool and the role it plays in the operation. Frequently mentioned in the context of the disturbance caused by a drone, its justified use, or what was unexpected about the operation. | "I also thought, (...) that it was not marked, not somehow in the police colours or something, that it now belonged to the police. It could have been a private person. That's why I didn't see it as security, it could also have been someone who is just media-hungry." (D) |

| drone use value (15) | Statements referring to value contributed by the drone to the general society, bystanders, or witnesses. Frequently an additional explanation or discussion sparked by the question on justified use of drones when the participants refer to the value of drone use as prerequisite for justification of its employment. | "I didn't feel it was unpleasant. (...) Because I trust that the drone helps to catch the the robbers faster." (F) "Ich denke auch, es ist ein Teil der Arbeitsausrüstung von der Polizei, was sicher hilft, um die Umgebung schneller abzusichern, weil sie von oben alles sehen sozusagen." (D) |
|---|---|---|
| transparency (48) | Statements referring to the sense of transparency concerning the use of drones by the police or the operation. Differs from the recognizability such that it refers to information that could be provided by humans rather than indicated by the drone or its design. | "I know in the future that the police have drones, today I don't think they have any. Then I will also know what is going on. Ah, that's probably the police, then it wouldn't worry me. So, you would certainly have to inform the people. But that would be a big issue anyway. Unless it has already been introduced and we don't know." (G) |
| additional considerations (115) | Other considerations not mentioned in other codes and not directly addressed by the questions. This code was split in the subsequent stage of the analysis contributing to multiple categories presented in Table 4, sometimes combined with segments from other codes. | "I was simply wondering (...) in an emergency, if something happens, the drone is allowed to land on its own, that there is still trust in the technology. But then not even when flying around. I found this answer quite interesting. And that would interest me now." (E) |

## Group Discussions Guide

The following questions were used to guide the group discussions. They were primarily employed as prompts to cover the interesting topics. However, the focus of the conversations and the order of questions was adjusted to the themes appearing in the conversations.

1) How did you experience the police operation? Was there anything unexpected? How did it impact your perceptions?

2) What impact did the drone have on your feeling of safety? Why?

3) Did you feel disturbed by the drone? Why?

4) If you had to summarise in one sentence: Who (or what) is the drone for you? Why?

5) Do you think the drone is capable of doing good/right or wrong in principle? What would be an example of right or wrong for you?

6) Who is responsible or to whom is it due if the drone does right or wrong? Why?

7) Under what circumstances would you consider the use of the drone justified? Why?

8) Do you have any other comments?